\documentclass{emulateapj}
\usepackage{threeparttable,textcomp}
\usepackage{tensor}

\usepackage{color}

\shorttitle{Mg{\sc ii} absorbers towards GRBs}
\shortauthors{Rapoport et al.}

\begin{document}

\title{On the significance of the excess number of strong Mg{\sc ii}
  absorbers observed towards Gamma-ray bursts}

\author{Sharon Rapoport\altaffilmark{1}, Christopher A. Onken\altaffilmark{1}, J. Stuart B. Wyithe\altaffilmark{2} ,Brian
  P. Schmidt\altaffilmark{1}, Anders O. Thygesen\altaffilmark{3}}

\altaffiltext{1}{Research School of Astronomy and Astrophysics, The Australian
  National University, Canberra 2611, Australia}
\altaffiltext{2}{School of Physics, University of Melbourne,
  Parkville, Victoria 3010, Australia}
\altaffiltext{3}{Zentrum f\"{u}r Astronomie der Universit\"{a}t Heidelberg,
Landessternwarte, K\"{o}nigstuhl 12, 69117 Heidelberg, Germany.}
\begin{abstract}
  The number of strong (equivalent width $>1$\AA) Mg{\sc ii} absorbers
  observed towards Gamma-ray bursts (GRBs) has been found to be
  statistically larger than the number of strong absorbers towards
  quasi-stellar objects (QSOs). We formalise this ``Mg{\sc ii}
  problem'' and present a detailed explanation of the statistical
  tools required to assess the significance of the discrepancy. We
  find the problem exists at the $4\sigma$ level for GRBs with
  high-resolution spectra. It has been suggested that the discrepancy
  can be resolved by the combination of a dust obscuration bias
  towards QSOs, and a strong gravitational lensing bias towards
  GRBs. We investigate one of the two most probable lensed GRBs that
  we presented in our previous work (GRB020405;
  \citeauthor{Rapoport2012}) and find it not to be strongly
  gravitationally lensed, constraining the percentage of lensed GRBs
  to be $<35\%$ (2$\sigma$). Dust obscuration of QSOs has been
  estimated to be a significant effect with dusty Mg{\sc ii} systems
  removing $\sim$20\% of absorbed objects from flux-limited QSO
  samples. We find that if $\sim$30\% of the strong Mg{\sc ii} systems
  towards QSOs are missing from the observed samples, then GRBs and
  QSOs would have comparable numbers of absorbers per unit
  redshift. Thus, gravitational lensing bias is likely to make only a
  modest contribution to solving the Mg{\sc ii} problem. However, if
  the dust obscuration bias has been slightly underestimated, the
  Mg{\sc ii} problem would no longer persist.
\end{abstract}


\section{Introduction}
The number of strong Mg{\sc ii} absorbers per unit redshift along Gamma-ray burst (GRB)
lines-of-sight (LOS) has been found to be statistically different than
the rate towards quasi-stellar objects (QSOs).  \citeauthor{Prochter2006}
(\citeyear{Prochter2006}, hereafter P06) found 14 strong absorbers
along the LOS to 14 GRBs, and \citeauthor{Vergani2009}
(\citeyear{Vergani2009}, hereafter V09) increased the sample
and found 22 strong absorbers along the LOS to 26 GRBs. In contrast,
studies of QSOs have found Mg{\sc ii} absorbers in only $\sim25\%$ of
sight lines \citep{Prochter2006b}. Both being high-redshift beacons,
GRBs and QSOs are expected to have similar LOS through the cosmos, and
explaining the preponderance of Mg{\sc ii} absorbers towards GRBs has
proven a challenge.

In the effort to unveil this problem, \cite{Porciani2007} considered
dust obscuration, beam size differences, intrinsic properties of GRBs
and gravitational lensing as possible causes. They found beam size
differences irrelevant, with simulations predicting the absorbing
systems are significantly larger than either
beam. \cite{Cucchiara2009} found no dissimilarities between the
populations of absorbers, suggesting it is unlikely that there are
excess absorbers physically associated with the GRBs. This leaves dust
obscuration of QSOs and gravitational lensing of GRBs as the two most
plausible explanations.

\citeauthor{Wyithe2011} (\citeyear{Wyithe2011}; hereafter
W11) suggested that, being detected in two independent energy bands,
GRBs could be subjected to a multi-band magnification bias (see
\citeauthor{Wyithe2003} \citeyear{Wyithe2003} for
details). Consequently, observed GRBs could be more likely to be strongly
gravitationally lensed than QSOs, with a resulting increase in
absorber numbers due to the lensing galaxies. Following this prediction, \cite{Rapoport2012}
investigated the probability distribution of alignment between GRBs
with strong Mg{\sc ii} absorption and the galaxy closest to their
LOS. The most interesting cases to suggest possible strong
gravitational lensing scenarios were those of GRB020405, which showed
a second transient $\sim3^{\prime\prime}$ from the GRB's optical
afterglow location, and GRB030429, which is aligned only 1.2\arcsec\
away from a massive galaxy that is known to have caused strong Mg{\sc ii} absorption.

As a part of their study of dusty systems along QSO LOS,
\citeauthor{Budzynski2011} (\citeyear{Budzynski2011}; hereafter BH11)
tested the bias in absorber numbers expected due to dust obscuration
by foreground objects towards QSOs and concluded that while some bias
is anticipated, it cannot solely account for the discrepancy. From the
total sample of QSOs they studied, they found that $24\%\pm4\%$ of
strong Mg{\sc ii} absorbers are not observed because the QSOs are
obscured and either fall below the typical signal-to-noise (S/N)
thresholds or are no longer detected.

In this paper we quantify the significance of the Mg{\sc ii} problem
when including 3 new LOS, and describe the statistical tools we use
for this purpose in Section \ref{sec:significance}. In Section
\ref{sec:bias}, we analyze the gravitational lensing bias, presenting
new observations of the lensing candidate, GRB020405; discuss the dust
obscuration bias; and calculate the total bias required to resolve the
discrepancy. Our conclusions are summarized and discussed in Section
\ref{sec:disc}.

\section{Significance of the Mg{\sc ii} Problem}\label{sec:significance}
To quantify the significance of the Mg{\sc ii} problem we require
knowledge of the redshift path over which strong systems can be
observed (also known as path density, $g(z)$) towards the GRBs. The
appropriate comparison for the number of observed systems towards GRBs
is the expected number of systems towards QSOs covering an equivalent
$g(z)$, which is calculated from the QSO absorber number density
($\partial n/\partial z$) as
\begin{equation}
\label{eq.NmgIIexp}
N^{\mathrm{MgII}}_{\mathrm{exp,QSO}}=\int g(z)\frac{\partial
n}{\partial z}dz.
\end{equation}

Recently, \citeauthor{Lawther2012}~(\citeyear{Lawther2012}; hereafter
L12) used SDSS DR7 to study QSOs with Mg{\sc ii} absorption lines. For
each redshift bin they measured how many QSO spectra would allow
detection of a strong Mg{\sc ii} system at that redshift, and the
number of absorbers $N(z)$ observed in that bin. They find that for
strong absorbers (equivalent width $(EW)>1$\AA), the number density
per co-moving Mpc along a LOS, $n(X)$, can be well described as a
function of redshift, $z$, by the following expression: 
\begin{equation}
cn(X)/H_{0} = n_{0}\exp(z_{0}/z),
\end{equation}
where $n_{0}=0.110\pm0.005$ absorbers per unit redshift, and
$z_{0}=-0.11\pm0.03$. The number density, $\partial n / \partial z$ is
given by:

\begin{equation}
\label{eq.n(z)}
\frac{\partial n}{\partial z}=\frac{cn(X)}{H_0}\frac{(1+z)^2}{\sqrt{\Omega_M(1+z)^3+\Omega_\Lambda}}.
\end{equation}

In 2006, P06 reported that the probability of finding the number of
strong Mg{\sc ii} absorbers observed towards 14 GRBs in a similar
redshift path towards QSOs was $<0.1\%$, using a Monte-Carlo (MC)
analysis. In 2009, V09 analysed all GRBs with high-resolution spectra
taken by UVES on the VLT up to June 2008 (10 objects). Their complete
sample, which included 26 GRBs (16 with available data from the
literature, including those from the P06 sample), revealed 22 strong
Mg{\sc ii} systems. They found that for a similar redshift path
towards QSOs, one would expect 10.41 strong absorbing
systems. However, they incorrectly included Poisson statistics to express
the error for this number when it was calculated from a fit to the
absorber number density equation ($\partial n/\partial z$) for
QSOs. We advocate instead that it is best to use a MC approach to
reflect the distribution of absorbers, such as was undertaken by P06.

Here, we revise the V09 work, deploying different statistical
tools, and also include 3 new GRBs with high-resolution VLT data
(GRB080804, GRB081008 and GRB081029). 

\subsection{Additional Data}
We reduced and analyzed publicly available data for GRB080804, which
we acquired from the ESO archive. The data were taken using the UVES
instrument on the VLT, with a total 2.6 hours of integration time. The
data were acquired with a 1\arcsec\ slit in good seeing
($\sim0\farcs7$). The observations were reduced using the ESO Gasgano
pipeline for
UVES\footnote{http://www.eso.org/projects/dfs/dfs-shared/web/vlt/vlt-instrument-pipelines.html}.
The pipeline performs the standard tasks of bias subtraction,
flat-fielding, spectral order location, wavelength calibration and
extraction of the spectrum. Candidate Mg{\sc ii} doublets were
identified by visual inspection of the spectra. The high signal to
noise ratio of the data allowed lines exhibiting EWs as low as 0.1\AA\
to be detected. All candidate doublets were then shifted to the
rest-frame to assess if the lines matched the Mg{\sc ii} doublet at
2794.4\AA ~and 2801.5\AA. All equivalent widths were measured in
IRAF\footnote{http://iraf.noao.edu} using the ``splot" task
\citep{Tody1993}. The total wavelength coverage is 3,600\AA \---
9,000\AA ~with telluric features preventing possible identification of
the doublet between 5,598\AA \--- 5,670\AA, 6,850\AA \--- 6,900\AA,
7,170\AA \--- 7,350\AA\ and 7,520\AA \--- 7,670\AA . The wavelength
limits were converted into redshift space for inclusion in the
redshift path density. No strong Mg{\sc ii} absorbing system was
identified.

The redshift path and absorber information for GRBs 081008 and 081029
were taken from the literature
\cite[respectively]{D'Elia2011,Holland2012}. The full dataset is
listed in Table~\ref{datasample}, and the corresponding redshift path
density is shown in Figure~\ref{fig:g_z}.

\subsection{Statistical Analysis}
We calculate the number of strong Mg{\sc ii} absorbers one would
expect to find towards QSOs for the redshift path of our observed GRBs
using equation \ref{eq.NmgIIexp}. Using the V09 redshift path density
for only the 10 UVES GRBs, which included 9 strong absorbing systems,
L12 predicted that $\tensor*{4.1}{^{+0.8}_{-0.7}}$ strong absorption
systems would be expected from a similar QSO sample. While the
relatively small errors represent the degree of precision in the mean
expected value for strong absorbers along QSOs LOS, we note that the
probability around the mean value is not a Gaussian distribution, and
therefore that the error cannot be used to directly estimate the
significance of the Mg{\sc ii} problem.

\begin{figure}
\begin{center}
\includegraphics[width=\columnwidth]{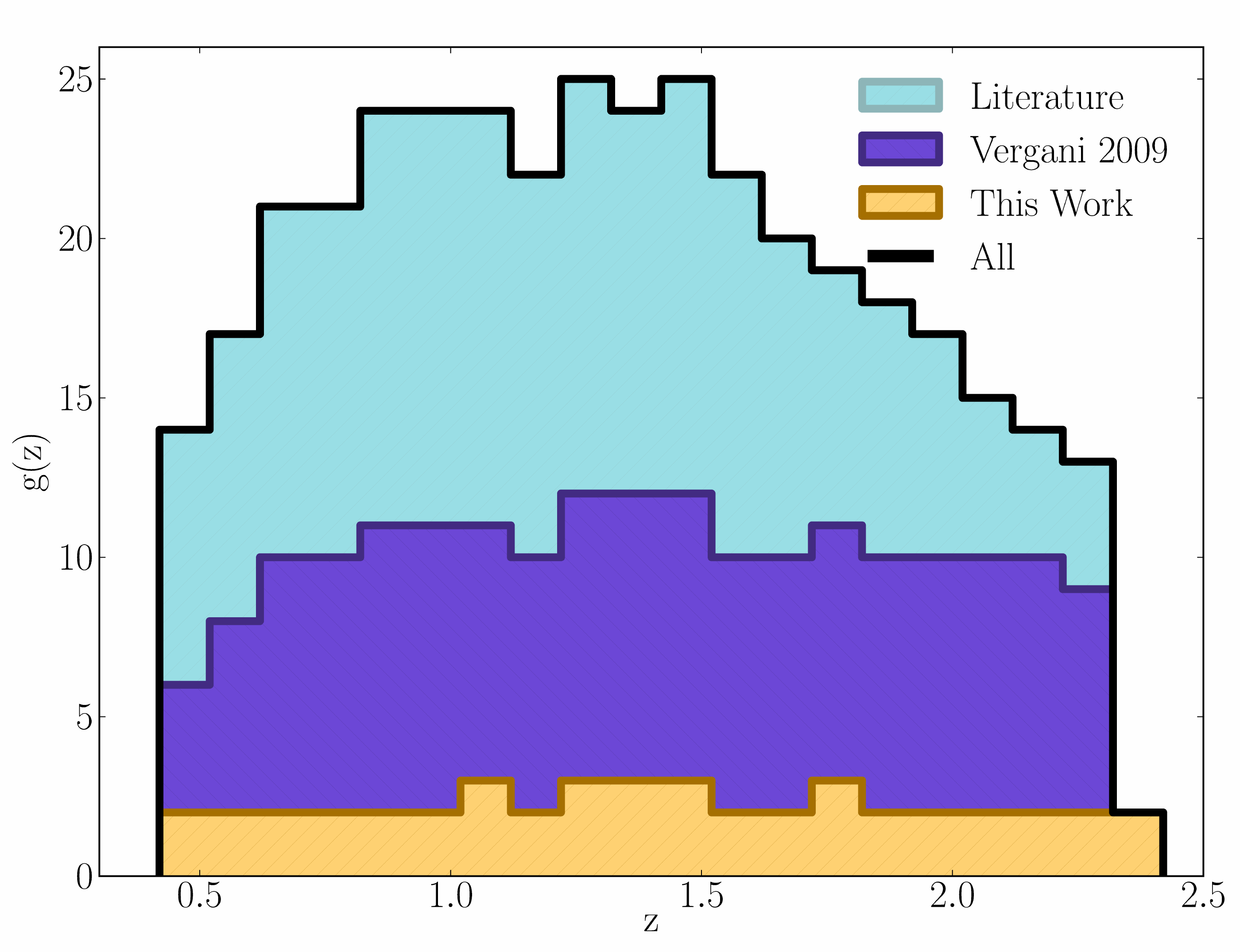}
\caption[short]{Redshift path density g(z) for absorbers with
  $EW>1$\AA~ towards the GRBs used in this study. The contribution
  labelled "Vergani 2009" includes only the 10 UVES GRBs, while the
  values from the literature were also included in the full sample in
  V09. The objects we added to the study are shown in
  yellow. The total sample used in the paper is the sum of all the
  different contributions and is outlined with the black curve.}
\label{fig:g_z}
\end{center}
\end{figure}

Using $\partial n / \partial z$ from L12, and the $g(z)$ for our 29
GRBs, we find $N^{\mathrm{MgII}}_{\mathrm{exp}}=9.6\pm 0.7$ for a
comparable sample of QSO LOS, where the errors are propagated from the
uncertainties of the fitting parameters for $\partial n / \partial z$
($n_0$ and $z_0$). When using the P06 values for $n_0$ and $z_0$
(values found for SDSS DR3),
$N^{\mathrm{MgII}}_{\mathrm{exp,QSO}}=7.4\pm0.7$ with the $g(z)$
provided in V09 only, and $8.3\pm 0.8$ for the $g(z)$ included in this
work. The difference between the GRB and the mean of the QSO
populations is reduced when using our sample relative to the earlier
V09 sample. It is further reduced when using the updated
L12 fit instead of the P06 fit to the QSO number density (as the
expected value for QSO absorbers increases). Thus, as more data are
available, both for GRBs and QSOs, the difference between the expected
number of absorbers towards QSOs and the observed population towards
GRBs is seen to decline, implying that we are still strongly affected
by small number statistics.

In order to assess the significance of the current Mg{\sc ii} problem we
calculate how unlikely it would be to observe 22 strong Mg{\sc ii}
systems along the LOS to 29 GRBs, under the hypothesis that GRBs and
QSOs probe the same absorber population. We use the SDSS DR4 catalog
of \cite{Quider2011}, which includes 44,600 QSOs spectra and contains
$\sim17,000$ measured Mg{\sc ii} doublets. The catalog provides
information regarding the redshift of the QSO and the redshift and EW
of the Mg{\sc ii} absorber. We conduct a MC simulation in which we
randomly select QSOs from the sample with similar $g(z)$ to the
observed GRBs, and count the number of strong absorbers towards
them. We are not constrained to match the precise number of source
systems, as the absorber rate depends only on the redshift path being
probed. Therefore, for each redshift bin of the $g(z)$ function (see
Figure~\ref{fig:g_z}) we select QSOs at a redshift of a randomly
selected GRB (within a 0.1 redshift bin), constraining the GRB sample
to those at a redshift which is larger than the bin, and small enough
so the tested redshift bin does not fall shortward of the Lyman alpha
break in the QSO spectrum. Once an appropriate random QSO is chosen,
we count the number of absorbers within the redshift bin. We continue
this procedure until we have covered the full redshift path of our
$g(z)$ function. Our MC analysis includes $10,000$ trials of the full
path selection.

Our analysis reproduces the mean value calculated above from the
empirical fits of P06 (8.3 absorbers). The histogram of the number of
absorbers is consistent with a Gaussian distribution, for which we
measure a standard deviation of $\sigma=3.2$ absorbers. Our simulation
resembles the results of P06 more than those of L12 as we are using
the QSO Mg{\sc ii} absorbers catalog from SDSS DR4. The SDSS DR4 is
the largest publicly available catalog which includes all the data
required to conduct this analysis. P06 found fewer absorbers per unit
redshift in the DR3 sample than the L12 DR7 study, as seen in Figure 3
of L12. The observation of 22 absorbers among a population with a mean
of 8.3 and $\sigma=3.2$ would be 3.9$\sigma$ from the average
value. As expected if the distribution is Gaussian, we find no
instances of a sample with 22 or more strong absorbers in our 10,000
MC trails. To test the sensitivity of our results to the chosen QSO
redshift we repeat the MC analysis and do not constrain the QSOs to
have similar redshifts as the GRBs. We do not find any statistically
significant deviations in the results, either in the absorber numbers
or in the redshift distribution of the absorbing systems. Moreover,
the redshifts of the absorbing systems are similar between the QSOs
and the GRBs.

\section{Potential Bias Contributions}\label{sec:bias}

Having found the difference between the GRB and QSO lines of sight to
be statistically significant, we now explore the two leading
explanations for this discrepancy: gravitational lensing bias and dust
obscuration bias.

\subsection{Gravitational Lensing Bias}
A potential contributor to the Mg{\sc ii} problem is that GRBs with
observed afterglows are more likely to be gravitationally lensed than
QSOs (see W11 and references therein). This is attributed to the
multi-band magnification bias, arising from GRBs being detected in two
bands ($\gamma$--rays and optical) with uncorrelated intrinsic fluxes
(see \citealt{Wyithe2003} for details). If GRBs are preferentially
lensed compared to QSOs, a higher rate of MgII absorption would be
expected towards GRBs from the gaseous halos surrounding the lensing
galaxies.

\subsubsection{GRB020405}
Following the W11 prediction that, if gravitational lensing bias is
the explanation for the Mg{\sc ii} problem, then $10-60\%$ of GRBs
with strong Mg{\sc ii} absorption should have been lensed strongly
enough to produce multiple images, we studied archival data of the
GRBs from the V09 sample to look for potential strongly
gravitationally lensed systems \citep{Rapoport2012}. Two potential
cases of lensing were identified, one of which was GRB020405
(z=0.695), which had a nearby ($3^{\prime \prime}$) transient that
could have been a repeating image of the same GRB.

\begin{figure}[htb]
\begin{center}
\includegraphics[width=\columnwidth]{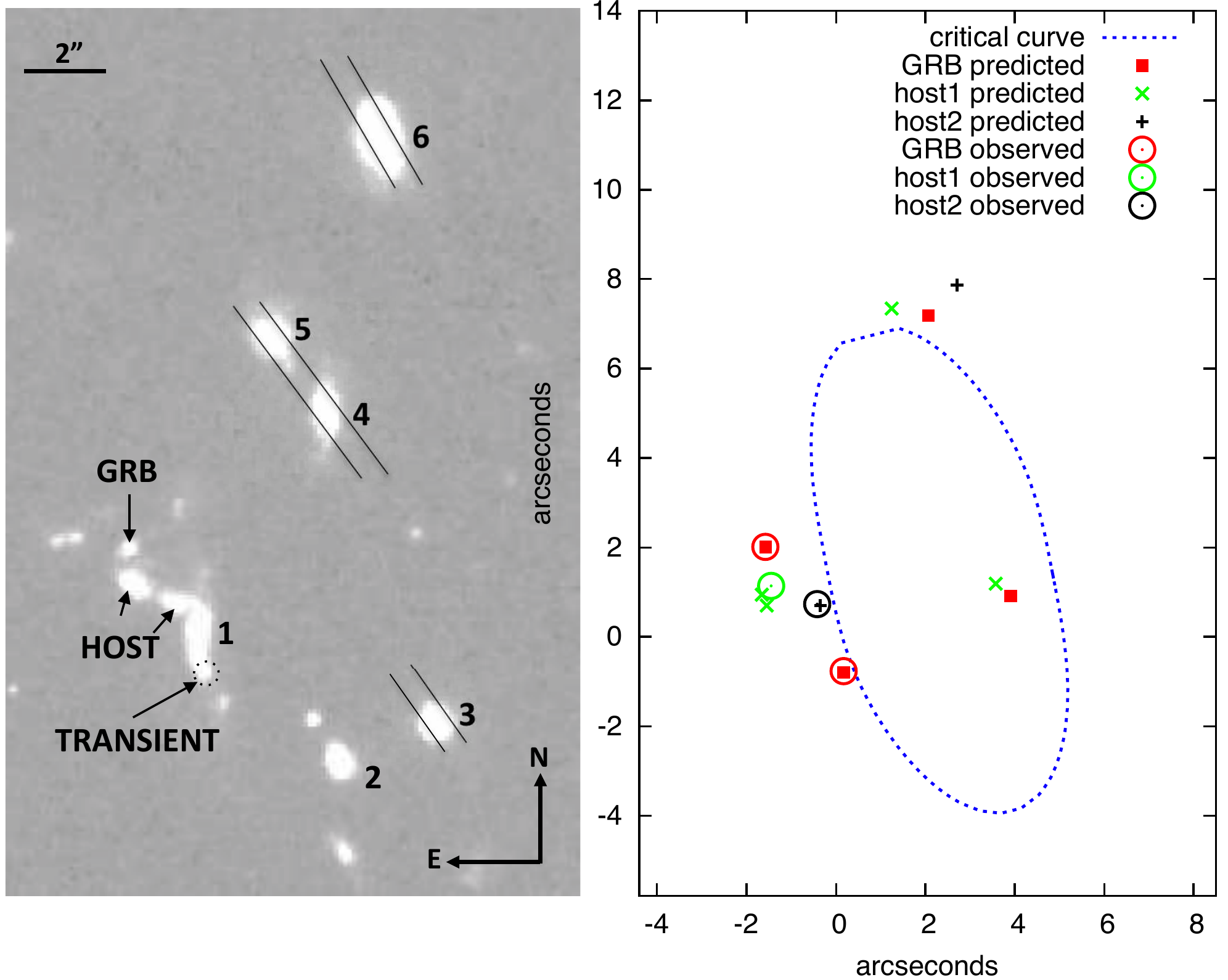}
\caption[short]{\emph{Left:} HST WFPC2/F702W field of GRB020405. The
  GRB is clearly visible and the complex host and second transient are
  indicated. Objects 1 and 2 were found to be at the strong absorber's
  redshift of z=0.472. Our observations indicate that object 3 is a
  field star, galaxy 4 is a galaxy at $z=0.485\pm0.002$, object 5 has
  unknown redshift and galaxy 6 is at $z=0.484\pm0.002$. The slits
  used in the Gemini/GMOS observations are superimposed on the
  image. \emph{Right:} LENSMODEL solution suggested in Rapoport et
  al. (2012), which assumed objects 1-6 were at the strong absorber
  redshift of z=0.472. With the newly measured redshifts of objects 4
  and 6, we are not able to identify any gravitational lensing model
  to explain the transient as another image of the GRB.}
\label{fig:grb020405model}
\end{center}
\end{figure}

The transient near GRB020405 was first observed by \cite{Masetti2003}
in an HST image taken $\sim19$ days after the \emph{Swift}
trigger. They noted that it was not visible in F555W but only in F702W
and F814W, and had completely faded by the last observation in August
2002. Figure~\ref{fig:grb020405model} shows the locations of the GRB,
transient and host galaxy. Objects 1 and 2 were confirmed, using VLT
spectroscopy, to be at the strong absorber redshift ($z=0.472$).
Using GRAVLENS \citep{Keeton2001} to model the system, we found that
if objects 3-6 (or even only 4-6) are part of a group at the
absorbers' redshift, the observations could be explained as a multiply
imaged GRB (see \citealt{Rapoport2012} for details and
Figure~\ref{fig:grb020405model} right panel for a potential model). In
order to test this scenario we initiated a detailed analysis of the
surrounding galaxies.

We obtained spectra for several of the galaxies around GRB020405 using
the GMOS instrument \citep{Hook2004} on Gemini-South (PI Rapoport,
Program ID GS-2012A-Q-9). The multi-slit data were taken on UT 03
April 2012 and consisted of eight 120s exposures with the R150
grating. The 1\arcsec\ slit widths provided a spectral resolution of
$R\sim 315$, and the data were binned 4$\times$4 to yield a spatial
scale of $\sim 0.3$\arcsec~pixel$^{-1}$ and a dispersion of
13.6\AA~pixel$^{-1}$. To improve the removal of sky and fringing
features, the mask was cut with two sets of slits, offset by
85\arcsec, and the field was dithered between the two positions. The
data were reduced using the standard Gemini/IRAF
packages\footnote{\url{http://www.gemini.edu/sciops/data-and-results/processing-software}}
for flat fielding and wavelength solution. No standard flux correction
was applied as the only information needed from the observations were
redshifts.

\begin{figure}[htb]
\begin{center}
\includegraphics[width=\columnwidth]{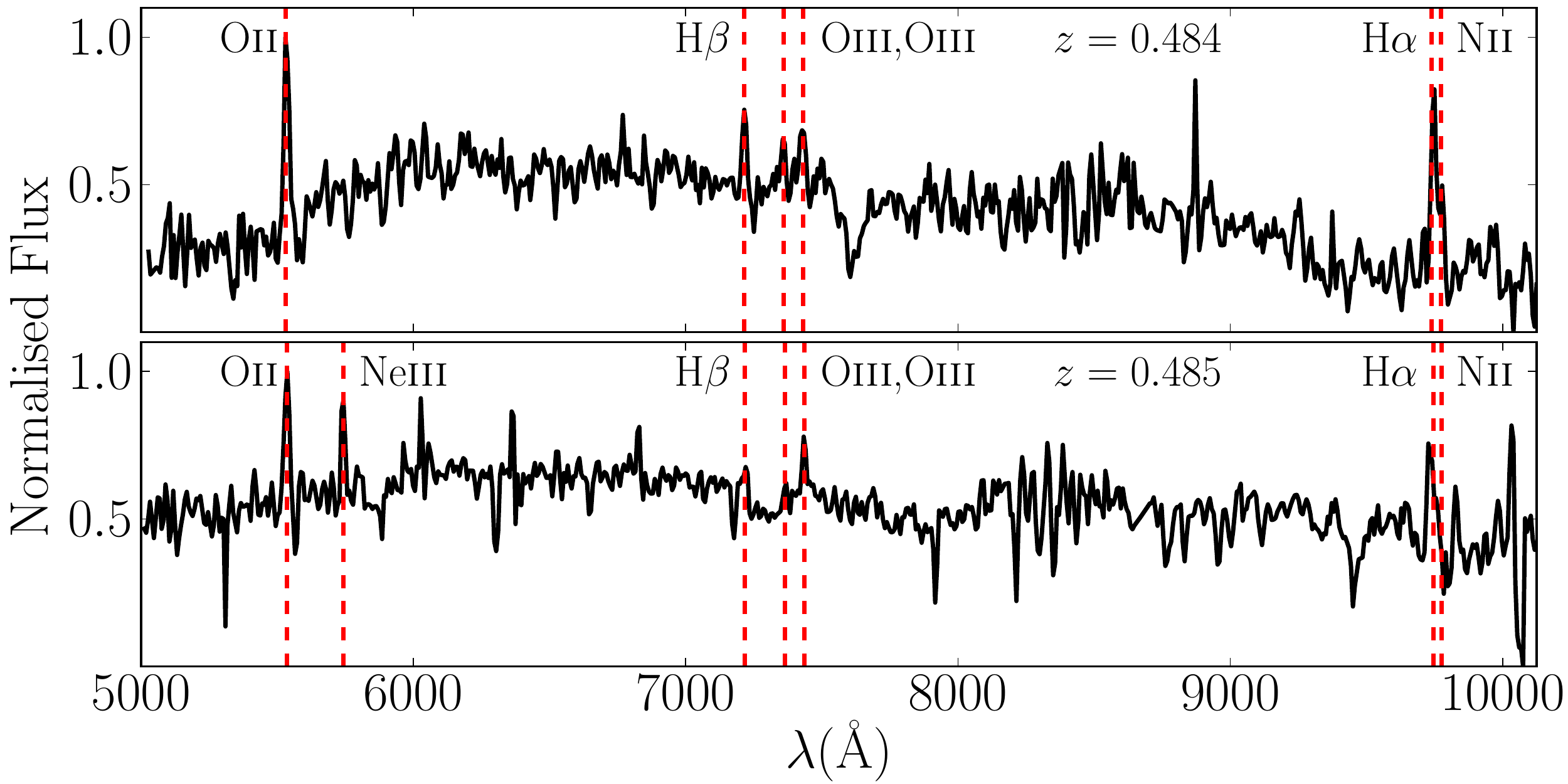}
\caption[short]{GMOS spectra of galaxies 4 (\emph{lower} panel) and 6
  (\emph{upper} panel), nearby galaxies to GRB020405 (see
  Figure~\ref{fig:grb020405model} for alignment). The strongest emission lines
  are identified and find galaxy 4 to be at $z=0.485\pm0.002$ and galaxy 6 at $z=0.484\pm0.002$.}
\label{fig:spectra}
\end{center}
\end{figure}

Comparing our reduced spectra to a library of
templates using the RUNZ 2dFGRS redshift code (R. Sharp 2012, private
communication, see Figure~\ref{fig:spectra}), we were able to
determine that objects 4 and 6 are galaxies at $z=0.485\pm0.002$ and
$z=0.484\pm0.002$ respectively, where the errors on the redshifts are
based on a centroid fit to the strongest line. Object 3 is consistent
with being a foreground star.  The spectra for object 5 were not of
sufficient signal-to-noise to determine the type or redshift of the
object.

As the galaxies at $z=0.472$ and $z=0.484$ are too distant to be from
the same halo ($\Delta v=4800$ km/s), but objects 4 and 6 are at a
similar redshift, we modelled the field using two different groups
(objects 1 and 2, and objects 4 and 6). In order to simplify the
mathematics we approximated the two groups (each as a singular
isothermal ellipse with shear) to reside in one plane at the strong
absorber redshift ($z=0.472$). No model was found which could explain
the observables using these galaxy redshifts and allowing plausible
masses for the galaxies.

Therefore, we conclude that the transient near GRB020405 was not
another image of that GRB, leaving one potential candidate as a strongly
gravitationally lensed GRB from the sample studied in our previous
work, GRB030429. This GRB was detected $1.2^{\prime\prime}$
away from a strong Mg{\sc ii} absorbing spiral galaxy. Using
photometric observations, a spectral energy distribution fitting
technique implied $M_B=-21.1\pm 0.1$ for the absorbing galaxy. The Tully-Fisher
relation for such an intrinsically bright galaxy suggests a velocity
dispersion of $160\pm65$ km/s, where $200$ km/s is required for
producing a second image for this GRB (see \citealt{Rapoport2012} for
further details). Thus, GRB030429 is likely to be magnified at some
level, possibly strongly. Verifying this scenario would require
measuring the velocity dispersion of the galaxy, which with
$R=22.70\pm0.12$ mag \citep{Jakobsson2004} and at a redshift of 0.841 would
be observationally very expensive. 

In our previous analysis of the existing HST imaging data for 11 GRBs
with Mg{\sc ii} absorption from the V09 sample, we ruled out strong
gravitational lensing in 6 cases. Now having determined that GRB020405
was not strongly lensed, we can place an upper (2$\sigma$) limit on
the fraction of strong absorbers which lead to multiply imaged GRB of
$\lesssim35\%$, though this is a very conservative limit as the
remaining systems show no evidence of lensing. Assuming only the most
likely case of GRB030429 is a viable strongly gravitationally lensed
candidate, the lensing fraction is reduced to $\lesssim13\%$ (2$\sigma$).

\subsubsection{Statistical Analysis}
To test the effect of strong gravitational lensing on the Mg{\sc ii}
absorber statistics, we account for the fraction of strongly
gravitationally lensed GRBs ($F_{\rm{lens}}$) by multiplying the
number of absorbers found in the MC trials by 1/(1-$F_{\rm
  lens}$). This approach assumes that all strongly gravitationally
lensed objects would show strong Mg{\sc ii} absorption. We find the
probability of finding 22 absorbers for a lensing fraction of 0.05
(assuming the Mg{\sc ii} absorber towards GRB030429 was the only one
out of the 22 systems that caused strong lensing) to be
0.02\%. Moreover, even if all 4 remaining candidates from
\cite{Rapoport2012} were strongly gravitationally lensed, the
probability of observing 22 absorbers would only increase to
0.7\%. This indicates lensing alone cannot account for the Mg{\sc ii }
problem.

\subsection{Dust Obscuration Bias}
The MgII problem could also arise by flux-limited QSO samples
selecting against absorbed objects with high MgII EW. Among a QSO
sample having the same redshift path density as GRBs in the V09
sample, BH11 showed that $\sim20\%$ (see Figure~19 in BH11) of sources
with strong absorbing systems would have been obscured by dust and so
either not detected or below the S/N threshold (for
$EW>1$\AA)\footnote{At higher redshift one would expect the QSOs to
  appear fainter and be more susceptible to dust obscuration. As the
  added $g(z)$ in this work, relative to that in V09, is flat (see
  Figure~\ref{fig:g_z}), the redshift path could be thought as being
  more concentrated at higher redshifts than it is in the V09
  sample. Therefore, by using the value found by BH11 for the
  percentage of obscured systems, we are more likely to be
  underestimating the number of obscured QSOs than vice-versa.}. Thus,
the QSO sample is likely missing $20\%$ of the strong absorbing
systems. To correct for this dust obscuration, we multiply the
predicted absorber rate from our MC analysis by a factor of
$1/(1-0.2)$. This correction results in a mean of 10.8 absorbers, with
a probability of finding 22 absorbers of 0.4\%.

\begin{figure}[htb!]
\begin{center}
\includegraphics[width=\columnwidth]{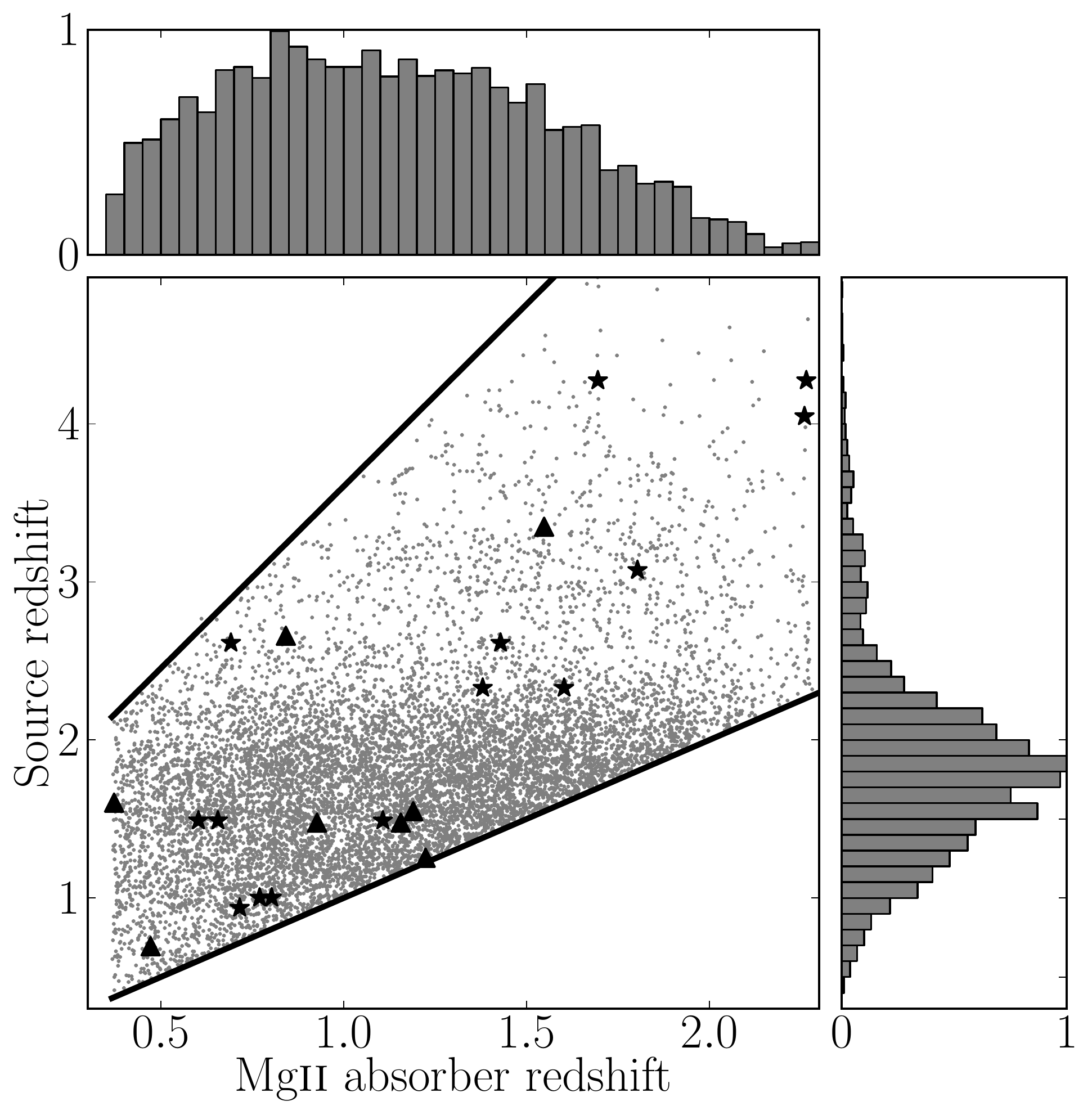}
\caption[short]{Redshifts of the QSOs, GRBs and their strong absorbing
  systems. The \cite{Quider2011} QSO sample used in the MC
  analysis is represented by grey points and their corresponding
  normalised histograms. The black triangles and the black stars
  mark the GRBs with absorbers from the full sample used in this
  paper, where the stars differentiate those GRBs which would
  have been excluded in the dust obscuration analysis of
  BH11. The lower black line represents the
  $z_{MgII}=z_{QSO}$ boundary and the upper black line is the limit
  due to Lyman alpha absorption.}
\label{fig:zqso_zabs}
\end{center}
\end{figure}

\subsection{Total Bias}
\begin{figure}[htb]
\begin{center}
\includegraphics[width=\columnwidth]{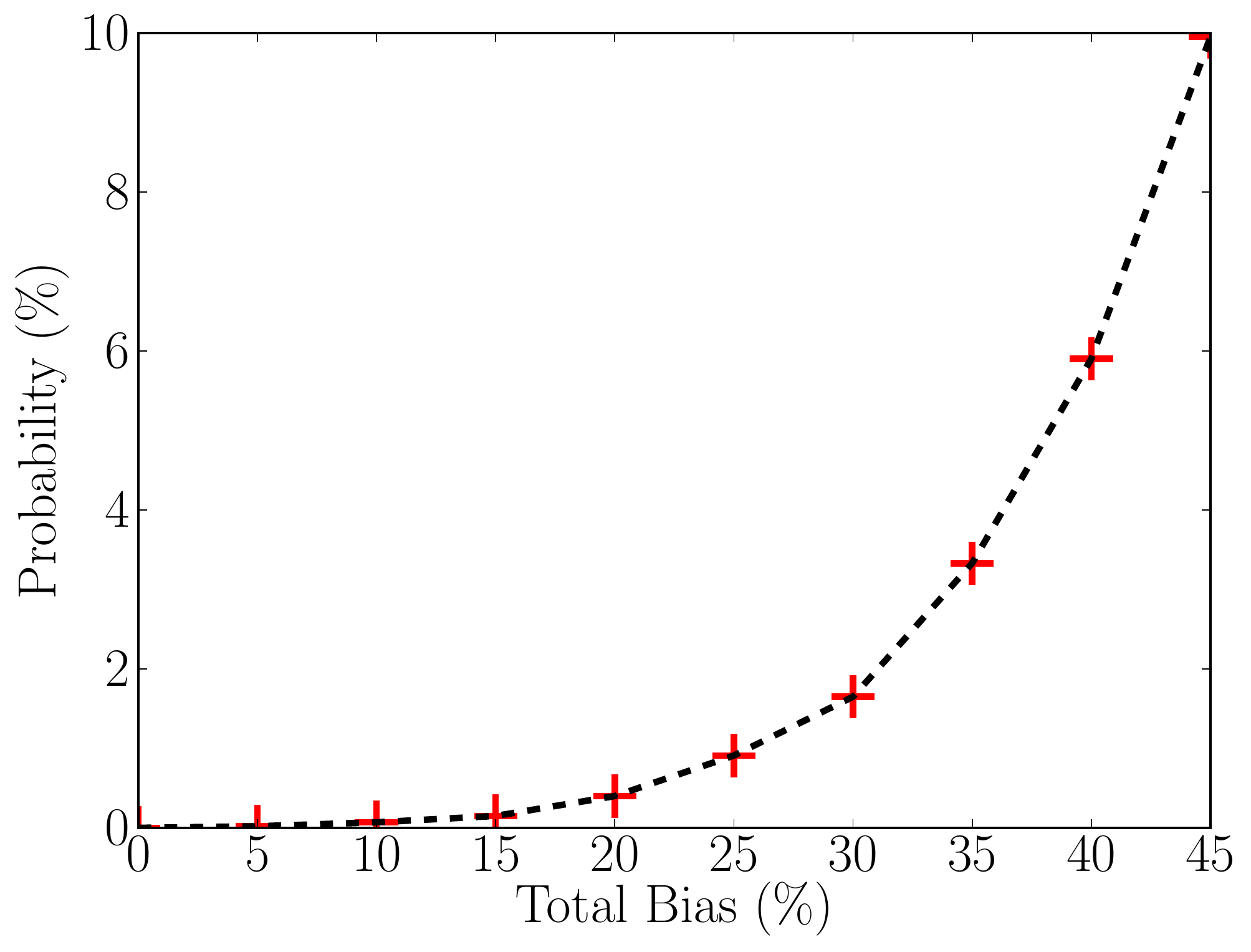}
\caption[short]{Probability of finding 22 strong absorbing systems
  along the LOS to QSOs having a similar redshift path density to that
  of the GRB sample for different fractions of total bias. The number
  of absorbers found along the QSOs LOS in the MC is multiplied by
  1/(1-$F_{\rm{Bias}}$) to account for the missing fraction before
  being compared to the number of absorbers found towards GRBs.}
\label{fig:flens}
\end{center}
\end{figure}

As neither dust obscuration nor gravitational lensing can by
themselves explain the Mg{\sc ii} problem, we repeat the MC
simulation and investigate the total bias required to resolve the
difference. We do this by imposing a bias on the QSOs by multiplying
the number of absorbers by $1/(1-F_{\rm{Bias}})$. We
then increment the bias fraction ($F_{\rm{Bias}}$) until it can
account for the observed excess towards GRBs. We find that for a total
bias fraction of $\sim30\%(45\%)$ the significance of the problem is
reduced to $2\sigma$ ($1\sigma$) (see Figure~\ref{fig:flens}). This
formalism assumes that the majority of the bias arises from the QSOs,
which is consistent with the $\sim20\%$ bias estimated by BH11.

BH11 noted that because their sample is flux limited they are likely
to be underestimating the fraction of absorbers missing from the SDSS
DR7, with systems having EW $>5.0$\AA~being completely
removed. Moreover, in their analysis they only use QSOs up to a
redshift of 3.5 and limit the number of absorbers per LOS to
two. Figure ~\ref{fig:zqso_zabs} shows the QSO, GRB and absorber
redshifts used in our MC analysis. To demonstrate the effect BH11's QSOs
selection would have on the GRB sample, we flag the GRBs that would
have been omitted with the star-like markers. The large fraction (60\%)
of absorbers along the LOS to GRBs which would have statistically gone
through more dusty systems than the QSOs that were examined by BH11
illustrates that the dust obscuration bias could be
underestimated. Therefore, while strong gravitational lensing might be
playing a minor role, the likelihood that the dust obscuration bias
has been underestimated implies that this effect can more easily
resolve the Mg{\sc ii} problem.

\begin{deluxetable}{llllc}
\tablecaption{GRB sample\label{datasample}}
\tabletypesize{\scriptsize}
\tablewidth{8.5cm}
 \tablehead{
 \colhead{GRB} &  \colhead{$z_{\rm{GRB}}$\tablenotemark{1}} & \colhead{$\Delta z$\tablenotemark{2}}&\colhead{$z_{\rm{abs}}$\tablenotemark{3}}&\colhead{Reference\tablenotemark{4}}}
 \startdata
991216&1.022&0.636&0.770&1\\
&&&0.803&\\
000926&2.038&1.392&\nodata&2\\
010222&1.477&1.022&0.927&3\\
&&&1.156&\\
011211&2.142&1.566&\nodata&1\\
020405&0.695&0.312&0.472&4\\
020813&1.255&0.866&1.224&5\\
 021004&2.3295&1.756&1.3800&6\\
 &&&1.6026\\
030226&1.986&1.590&\nodata&7\\
030323&3.372&0.822&\nodata&8\\
030328&1.522&1.131&\nodata&9\\
030429&2.66&1.241&0.8418&10\\
050505&4.275&0.856&1.695&11\\
&&&2.265&\\

 050730&3.9687&1.298&\nodata&6\\
 050820A&2.6147&1.845&0.6915&6\\
 &&&1.4288\\
050908&3.35&1.456&1.548&12\\
050922C&2.1996&1.682&\nodata&6\\
051111&1.55&1.036&1.19&12\\
060206&4.048&1.060&2.26&13,14\\
060418&1.4900&1.265&0.6026&6\\
&&&0.6559\\
&&&1.1070\\
060526&3.221&1.434&\nodata&15\\
060607A&3.0748&1.713&1.8033&6\\
071003&1.604&1.212&0.372&16\\
071031&2.6922&1.789&\nodata&6\\
080310&2.4272&1.841&\nodata&6\\
080319B&0.9378&0.57&0.7154&6\\
080413A&2.4346&1.650&\nodata&6\\
080804&2.20&1.63&\nodata&This work\\
081008&1.286&0.92&\nodata&17\\
081029&3.8479&1.50&\nodata&18
 \enddata
\tablenotetext{1}{Redshift of GRB}
\tablenotetext{2}{Redshift path length for strong Mg{\sc ii}
  absorbers}
\tablenotetext{3}{Strong absorber redshift}
\tablenotetext{4}{References 1: \cite{Vreeswijk2006} 2: \cite{Castro2003}
  3: \cite{Mirabal2002} 4: \cite{Masetti2003} 5: \cite{Barth2003} 6:
  \cite{Vergani2009} 7: \cite{Klose2004} 8: \cite{Vreeswijk2004} 9:
  \cite{Maiorano2006} 10: \cite{Jakobsson2004} 11: \cite{Berger2006}
  12: \cite{Prochter2006} 13: \cite{Chen2009} 14: \cite{Hao2007} 15:
  \cite{Thone2008} 16: \cite{Perley2008} 17: \cite{D'Elia2011} 18: \cite{Holland2012}}
\end{deluxetable}

\section{Discussion}\label{sec:disc}

We have presented an updated analysis of the GRB strong Mg{\sc ii}
absorber problem, including additional GRB lines of sight.
We outline a statistical method with which the Mg{\sc ii} problem
can be addressed, using a MC technique to estimate the number of
absorbers expected towards QSOs for an observed redshift path towards
GRBs. We find that the observed discrepancy between the number of strong
Mg{\sc ii} absorbers in GRBs and QSOs has a level of $4\sigma$
significance, with $<0.01\%$ chance of finding the observed number of absorbers
along GRB lines of sight in the MC results for the QSOs. 

Gravitational lensing bias for GRBs can contribute to the discrepancy,
and we examined the best lensing candidate of \cite{Rapoport2012},
GRB020405, which has a known foreground absorber and a second nearby
transient. We find that the nearby galaxies, which would need to be at
the absorber's redshift to allow possible strong gravitational lensing
by a galaxy group, are not at the appropriate redshift. With no
feasible lensing model found for the GRB and its nearby transient, we
conclude that GRB020405 was not affected by strong lensing. However,
there remains the possibility that gravitational lensing is playing a
role, as evidenced by the very close alignment of GRB030429 to a
massive foreground galaxy. If this GRB is significantly lensed, then
the Mg{\sc ii} discrepancy is slightly reduced, with the probability
of finding the number of strong absorbers along the GRBs increasing to
0.02\%. If 4 GRBs from the sample studied by \cite{Rapoport2012} are
strongly lensed, this implies a lensing fraction of 20\%, which would
increase the probability of finding the observed number of absorbers
in a comparable QSO sample to only 0.7\%. Thus we conclude that
lensing is very unlikely to solve the Mg{\sc ii} problem on its own.

The significance of the problem is also reduced after accounting for dust
obscuration bias, with the probability of finding the observed number
of absorbers along the LOS to the GRBs increasing to 0.4\%. When
including both dust and gravitational lensing biases, we calculate a
probability of 0.9\%, which remains statistically significant. In order
to reduce the problem to below a $2\sigma$ level, a total bias of
$\sim30\%$ is required. We suggest this could be largely satisfied by
the dust obscuration bias towards QSOs currently being underestimated.

Following submission of our paper, \cite{Cucchiara2012} increased the
sample of GRBs to 118 LOS by including GRBs with low-resolution
spectra. This study has found the rate of Mg{\sc ii} absorbers to GRBs
to be lower than previous studies, reducing the overall discrepancy
between the absorber rate to GRBs and QSOs to $<90\%$ confidence
level. If additionally, one were to include the dust obscuration bias
described above, we expect the rate of absorbers between GRBs and QSOs
to be completely consistent.

\section*{Acknowledgements}
SR gratefully acknowledges Robert Sharp for assistance with the
redshift identification of the Gemini galaxies, Stuart Sim for useful
conversation and the anonymous referee for invaluable comments. BPS
acknowledges financial support through an ARC Laureate Fellowship Grant
FL0992131.  JSBW acknowledges financial support through an ARC Laureate
Fellowship.  AOT acknowledges support from Sonderforschungsbereich SFB
881 ``The Milky Way System" (subproject A5) of the German Research
Foundation (DFG).  Based on observations obtained at the Gemini
Observatory, which is operated by the Association of Universities for
Research in Astronomy, Inc., under a cooperative agreement with the
NSF on behalf of the Gemini partnership: the National Science
Foundation (United States), the Science and Technology Facilities
Council (United Kingdom), the National Research Council (Canada),
CONICYT (Chile), the Australian Research Council (Australia),
Minist\'{e}rio da Ci\^{e}ncia, Tecnologia e Inova\c{c}\~{a}o (Brazil)
and Ministerio de Ciencia, Tecnolog\'{i}a e Innovaci\'{o}n Productiva
(Argentina)

\bibliographystyle{hapj}
\bibliography{grb020405}

\end{document}